\documentclass[a4paper]{article}

\usepackage{graphicx}\usepackage{multirow}\usepackage{amsmath,amssymb,amsfonts}\usepackage[title]{appendix}\usepackage{xcolor}\usepackage{textcomp}\usepackage{manyfoot}\usepackage{booktabs}\usepackage{listings}

\usepackage{orcidlink} 

\usepackage{authblk}

\date{}
\usepackage{hyperref}
\hypersetup{hidelinks=true}

\newcommand{\fnm}[1]{#1}
\newcommand{\sur}{}
\newcommand{\email}[1]{(\href{mailto:#1}{\tt #1})}
\newcommand{\orgdiv}{}
\newcommand{\orgname}{}
\newcommand{\orgaddress}{}
\newcommand{\city}{}
\newcommand{\state}{}
\newcommand{\country}{}
\newcommand{\keywords}[1]{}\newcommand{\backmatter}{}
\newcommand{\bmhead}[1]{\section*{#1}}
\newcommand{\bibcommenthead}{}

\usepackage[
    reversemp,
    paperwidth=210mm,
    paperheight=297mm,
    top={26mm},
    headheight={5.5pt},
    headsep={5.6mm},
    text={31pc,194.25mm},
    marginparsep=5mm,
    marginparwidth=12mm,
    bindingoffset=6mm,
    footskip=10mm]{geometry}

\usepackage{fancyhdr}
\fancypagestyle{firstpage}{
\fancyhf{}
\fancyfoot[L]{Notice: This manuscript has been authored by UT-Battelle, LLC, under contract DE-AC05-00OR22725 with the US Department of Energy (DOE). The US government retains and the publisher, by accepting the article for publication, acknowledges that the US government retains a nonexclusive, paid-up, irrevocable, worldwide license to publish or reproduce the published form of this manuscript, or allow others to do so, for US government purposes. DOE will provide public access to these results of federally sponsored research in accordance with the DOE Public Access Plan (\url{https://www.energy.gov/doe-public-access-plan}).}

}

\begin{document}
\thispagestyle{firstpage}

\title{Applying the FAIR Principles to computational workflows}

\author[*1]{
    \fnm{Sean R.} \sur{Wilkinson}~\orcidlink{0000-0002-1443-7479}
    \email{wilkinsonsr@ornl.gov}}
\affil[*]{corresponding author}
\affil[1]{
    \orgdiv{Oak Ridge Leadership Computing Facility},
    \orgname{Oak Ridge National Laboratory},
    \orgaddress{
        \city{Oak Ridge},
        \state{Tennessee},
        \country{USA}
    }
}

\author[2]{
    \fnm{Meznah} \sur{Aloqalaa}~\orcidlink{0000-0002-1382-3896}
    \email{meznah.aloqalaa@manchester.ac.uk}}
\affil[2]{
    \orgdiv{Department of Computer Science},
    \orgname{University of Manchester},
    \orgaddress{
        \city{Manchester},
        \country{UK}
    }
}

\author[3]{
    \fnm{Khalid} \sur{Belhajjame}~\orcidlink{0000-0001-6938-0820}
    \email{khalid.belhajjame@dauphine.fr}}
\affil[3]{
    \orgdiv{LAMSADE},
    \orgname{PSL, Paris Dauphine University},
    \orgaddress{
        \city{Paris},
        \country{France}
    }
}

\author[4]{
    \fnm{Michael R.} \sur{Crusoe}~\orcidlink{0000-0002-2961-9670}
    \email{crusoe@zib.de}}
\affil[4]{
    \orgdiv{Mathematics of Complex Systems division, Visual and Data-Centric Computing department, Bioinformatics in Medicine group},
    \orgname{Zuse Institute Berlin (ZIB)},
    \orgaddress{
        \city{Berlin},
        \country{Germany}
    }
}

\author[5]{
    \fnm{Bruno} \sur{de Paula Kinoshita}~\orcidlink{0000-0001-8250-4074}
    \email{bruno.depaulakinoshita@bsc.es}}
\affil[5]{
    \orgdiv{Earth Sciences},
    \orgname{Barcelona Supercomputing Center},
    \orgaddress{
        \city{Barcelona},
        \country{Spain}
    }
}

\author[6]{
    \fnm{Luiz} \sur{Gadelha}~\orcidlink{0000-0002-8122-9522}
    \email{luiz.gadelha@dkfz-heidelberg.de}}
\affil[6]{
    \orgdiv{German Human Genome-Phenome Archive (GHGA, W620)},
    \orgname{German Cancer Research Center (DKFZ)},
    \orgaddress{
        \city{Heidelberg},
        \country{Germany}
    }
}

\author[7]{
    \fnm{Daniel} \sur{Garijo}~\orcidlink{0000-0003-0454-7145}
    \email{daniel.garijo@upm.es}}
\affil[7]{
    \orgdiv{Ontology Engineering Group},
    \orgname{Universidad Politécnica de Madrid},
    \orgaddress{
        \city{Madrid},
        \country{Spain}
    }
}

\author[8]{
    \fnm{Ove Johan Ragnar} \sur{Gustafsson}~\orcidlink{0000-0002-2977-5032}
    \email{johan.gustafsson@unimelb.edu.au}}
\affil[8]{
    \orgdiv{Australian BioCommons},
    \orgname{University of Melbourne},
    \orgaddress{
        \city{Melbourne},
        \state{Victoria},
        \country{Australia}
    }
}

\author[2]{
    \fnm{Nick} \sur{Juty}~\orcidlink{0000-0002-2036-8350}
    \email{nick.juty@manchester.ac.uk}}

\author[9]{
    \fnm{Sehrish} \sur{Kanwal}~\orcidlink{0000-0002-5044-4692}
    \email{kanwals@unimelb.edu.au}}
\affil[9]{
    \orgdiv{Clinical Pathology},
    \orgname{University of Melbourne Centre for Cancer Research (UMCCR)},
    \orgaddress{
        \city{Parkville},
        \state{Victoria},
        \country{Australia}
    }
}

\author[8]{
    \fnm{Farah Zaib} \sur{Khan}~\orcidlink{0000-0002-6337-3037}
    \email{khanfarahzaib@gmail.com}}

\author[10]{
    \fnm{Johannes} \sur{Köster}~\orcidlink{0000-0001-9818-9320}
    \email{johannes.koester@uni-due.de}}
\affil[10]{
    \orgdiv{Bioinformatics and computational oncology (Bioinformatische Algorithmen in der Onkologie), Institute for AI in Medicine (IKIM), University Medicine Essen},
    \orgname{University of Duisburg-Essen},
    \orgaddress{
        \city{Essen},
        \country{Germany}
    }
}

\author[11]{
    \fnm{Karsten} \sur{Peters-von Gehlen}~\orcidlink{0000-0003-0158-2957}
    \email{peters@dkrz.de}}
\affil[11]{
    \orgdiv{Data Management Department},
    \orgname{Deutsches Klimarechenzentrum GmbH},
    \orgaddress{
        \city{Hamburg},
        \country{Germany}
    }
}

\author[12]{
    \fnm{Line} \sur{Pouchard}~\orcidlink{0000-0002-2120-6521}
    \email{lcpouch@sandia.gov}}
\affil[12]{
    \orgdiv{Center for Computing Research},
    \orgname{Sandia National Laboratories},
    \orgaddress{
        \city{Albuquerque},
        \state{New Mexico},
        \country{USA}
    }
}

\author[13]{
    \fnm{Randy K.} \sur{Rannow}~\orcidlink{0009-0005-7406-6170}
    \email{pi-boson@ieee.org}}
\affil[13]{
    \orgname{Silverdraft Supercomputing},
    \orgaddress{
        \city{Boise},
        \state{Idaho},
        \country{USA}
    }
}

\author[2,14]{
    \fnm{Stian} \sur{Soiland-Reyes}~\orcidlink{0000-0001-9842-9718}
    \email{soiland-reyes@manchester.ac.uk}}
\affil[14]{
    \orgdiv{Informatics Institute},
    \orgname{University of Amsterdam},
    \orgaddress{
        \city{Amsterdam},
        \country{The Netherlands}
    }
}

\author[15]{
    \fnm{Nicola} \sur{Soranzo}~\orcidlink{0000-0003-3627-5340}
    \email{nicola.soranzo@earlham.ac.uk}}
\affil[15]{
    \orgname{Earlham Institute},
    \orgaddress{
        \city{Norwich},
        \country{UK}
    }
}

\author[2]{
    \fnm{Shoaib} \sur{Sufi}~\orcidlink{0000-0001-6390-2616}
    \email{shoaib.sufi@manchester.ac.uk}}

\author[16]{
    \fnm{Ziheng} \sur{Sun}~\orcidlink{0000-0001-9810-0023}
    \email{zsun@gmu.edu}}
\affil[16]{
    \orgdiv{Center for Spatial Information Science and Systems, Department of Geography and Geoinformation Science},
    \orgname{George Mason University},
    \orgaddress{
        \city{Fairfax},
        \state{Virginia},
        \country{USA}
    }
}

\author[17]{
    \fnm{Baiba} \sur{Vilne}~\orcidlink{0000-0002-1084-7067}
    \email{baiba.vilne@rsu.lv}}
\affil[17]{
    \orgdiv{Bioinformatics Group},
    \orgname{Riga Stradins University},
    \orgaddress{
        \city{Riga},
        \country{Latvia}
    }
}

\author[18]{
    \fnm{Merridee A.} \sur{Wouters}~\orcidlink{0000-0002-2121-912X}
    \email{merridee.wouters@unsw.edu.au}}
\affil[18]{
    \orgdiv{School of Clinical Medicine},
    \orgname{University of New South Wales},
    \orgaddress{
        \city{Kensington},
        \state{New South Wales},
        \country{Australia}
    }
}

\author[19]{
    \fnm{Denis} \sur{Yuen}~\orcidlink{0000-0002-6130-1021}
    \email{denis.yuen@oicr.on.ca}}
\affil[19]{
    \orgname{Ontario Institute for Cancer Research},
    \orgaddress{
        \city{Toronto},
        \state{Ontario},
        \country{Canada}
    }
}

\author[2]{
    \fnm{Carole} \sur{Goble}~\orcidlink{0000-0003-1219-2137}
    \email{carole.goble@manchester.ac.uk}}

\maketitle

\abstract{Recent trends within computational and data sciences show an increasing recognition and adoption of computational workflows as tools for productivity and reproducibility that also democratize access to platforms and processing know-how.  As digital objects to be shared, discovered, and reused, computational workflows benefit from the FAIR principles, which stand for Findable, Accessible, Interoperable, and Reusable. The Workflows Community Initiative's FAIR Workflows Working Group (WCI-FW), a global and open community of researchers and developers working with computational workflows across disciplines and domains, has systematically addressed the application of both FAIR data and software principles to computational workflows. We present recommendations with commentary that reflects our discussions and justifies our choices and adaptations. These are offered to workflow users and authors, workflow management system developers, and providers of workflow services as guidelines for adoption and fodder for discussion. The FAIR recommendations for workflows that we propose in this paper will maximize their value as research assets and facilitate their adoption by the wider community.} 

\section*{Computational workflows and why FAIR matters}\label{sec:why-fair-matters}
Scale-ups in research data and the rise of data-driven science have spurred the research community to find solutions for managing and automating complex computational processes in order to ensure efficiency, reproducibility, scalability, collaboration, and quality-assured transparency. Computational workflows are a special kind of software specifically targeted at handling multi-step, multi-code data pipelines, data analyses, and other data-handling operations, especially through the efficient use of computational resources to transform data inputs into desired outputs.

Computational workflows are software with two prominent characteristics: (i) the composition (component number, order, structure) of \textbf{multiple components} that include other software, workflows, code snippets, tools, and services; and (ii) the \textbf{explicit abstraction} from the run mechanics in some form of \textbf{high-level workflow language} that precisely specifies the flow of data between the components, relying on a dedicated management system concerned with data handling and code execution. Metadata details the dependencies and computational requirements, including those of the computational environment \cite{SCHINTKE202482}.

At the simplest end of the complexity spectrum, the workflow language might consist of a script (e.g. Bash, Python, R) or enumerated stages in an electronic research notebook (e.g. Jupyter, RStudio, Apache Zeppelin, etc.) with a set of instructions that use inputs and outputs to pipe the results together. At the other end, workflows might employ a workflow management system (WMS) such as Nextflow \cite{DiTommaso2017}, Galaxy \cite{Galaxy2024}, Snakemake \cite{Molder2021}, or Parsl \cite{Babuji2019}. Using a WMS can provide a number of benefits including abstraction, scaling, automation, reproducibility \cite{kanwal-2017}, and provenance \cite{roadmap2022}. A fully featured WMS typically facilitates error handling and restarting, automatic data staging, provenance recording, handling of large datasets and complex computations, task and resource allocation, and distributed task execution to scale over local workstations, cloud resources, or high-performance computing clusters. WMSes and modular containerized software components (e.g. Docker, Singularity) aid portability and reproducibility, but they also face challenges (e.g., security issues when deployed in clusters, steep l earning curve, etc.). In practice, researchers use mixtures of techniques from across the spectrum, such as chaining scripts from a WMS that in turn is launched from a notebook.

\begin{figure}[ht]
\centering
\includegraphics[width=0.9\textwidth]{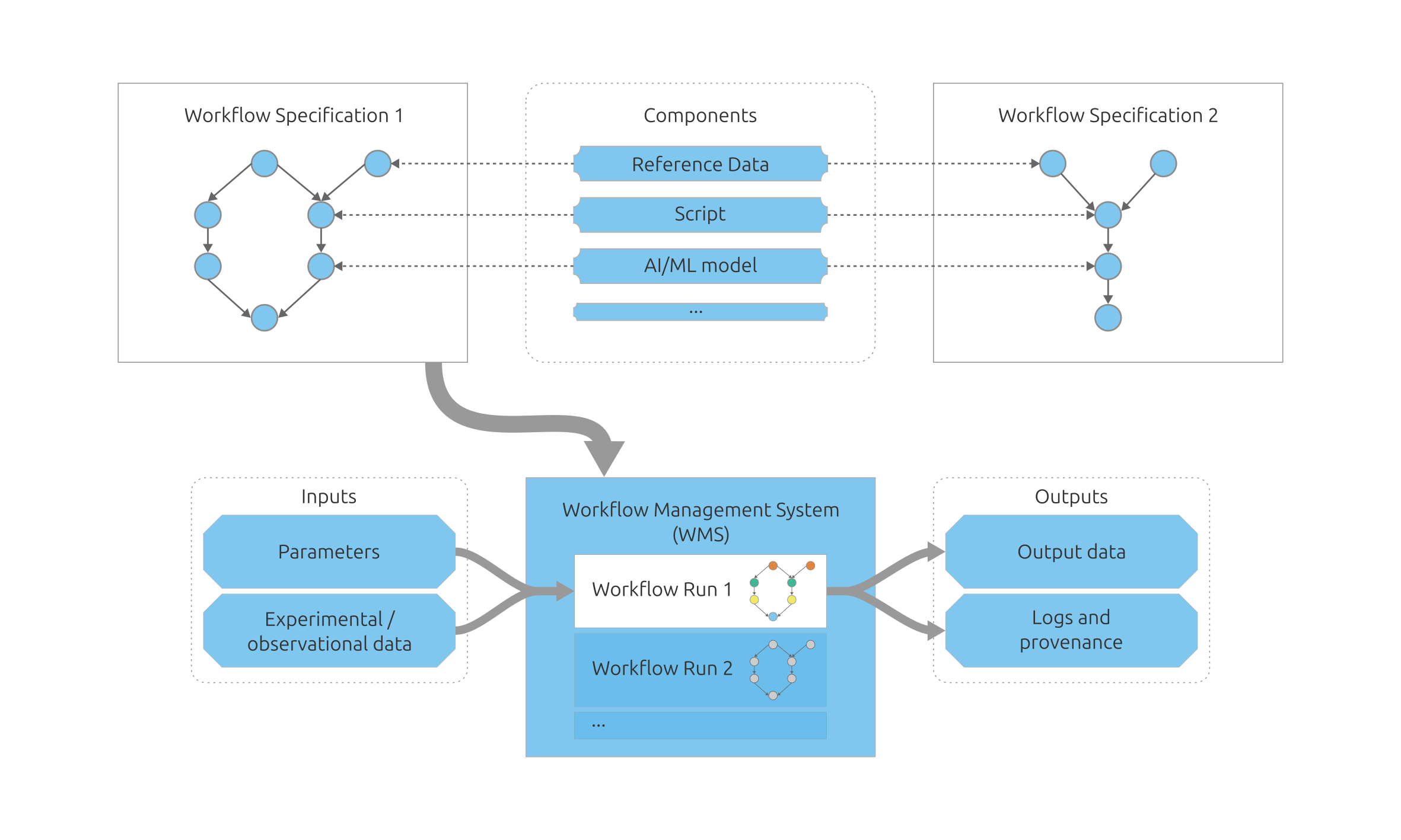}
\caption{A \textbf{workflow specification} formally specifies the data flow and/or execution control between \textbf{components} such as reference datasets, executable scripts, and AI/ML models. These components can be reused as parts of other workflow specifications. Using a workflow specification means instantiating it as a \textbf{workflow run} -- typically by executing it within a \textbf{workflow management system} -- by feeding it inputs like parameters and experimental/observational data as required. During execution, components will be used in some order, finally resulting in output data as well as logs and provenance metadata.}\label{fig:1}
\end{figure}

Workflows are complex digital objects with intra- and inter-object relationships that connect the entire workflow, its definition, and its scope. To help clarify our FAIR discussion, we present some definitions along with an illustration in Figure~\ref{fig:1}.

A \textbf{workflow} is the formal specification of the data flow and execution control between executable components, the expected datasets, and parameter files. Its complexity can be composed, and its overall process can be abstracted away from its execution. A \textbf{workflow run} is the instantiation of the workflow with inputs (parameters files, input datasets) and outputs (output data, the provenance execution log and lineage of data products).

A \textbf{workflow component} can be an executable and data. Executables include scripts, code, tools, containers, or workflows themselves remotely or locally executed, and native or third party; an example data component could be a reference dataset. For a workflow specification, data include metadata (e.g., its description), graphical images, test and benchmark datasets, parameter defaults, and so on. For a workflow run, data extends to the actual input and output datasets, the parameter files, and the provenance record detailing the code run and data product lineage. 

A \textbf{workflow management system} handles the data flow and/or execution control and performs the heavy lifting for computational and data handling.  A workflow management system abstracts the workflow from the underlying digital infrastructure, supporting scalability, portability, and differences in hardware architectures.

The use of workflows has accelerated in the past few years.  The existence of more than 350 different workflow management systems of varying maturity evidences the increasing popularity of workflows (\url{https://s.apache.org/existing-workflow-systems}). Workflows help reduce the burden of manual human effort by automating repetitive and time-consuming tasks. This automation enables efficiency in terms of resource allocation both by allowing humans to return to human-only tasks and by allowing the computational tasks to take advantage of opportunities for automatic optimizations. Workflows also automate critical processes that require standardization and increased reproducibility, improving efficiency by removing humans from the analytical process, at times dispensing with the need for manual management of data flow through a set of processing steps, and reducing the need for repeat analyses due to human error and/or bias. Moreover, automation ensures that computational experiments can be replicated. Workflows also provide a documented record of the computational processes, which helps in understanding, reviewing, and auditing those processes. As scientific activity often includes the exploration of analysis variance, modifying workflows to understand effects and changes on data products is simpler when those workflows are clearly described and comparable. Researchers can reuse workflows to re-analyze and adapt to incorporate new methods, tools, or data sources. Sharing and reusing workflows supports team collaboration and standardizes processes and analysis methods~\cite{garijo2014}.

The Findable, Accessible, Interoperable, and Reusable (FAIR) principles aim to maximize the value and impact of scientific digital objects. Such objects cannot be reused if they cannot be found, accessed, or understood; they cannot be combined if they are not interoperable. The FAIR guidelines for scientific data \cite{Wilkinson2016-vi} emphasize making data findable through metadata provision, unique and persistent identifier assignment, ensuring accessibility through open access or controlled access mechanisms, promoting interoperability through standardized formats and metadata schemas, and enabling reusability through clear licensing and documentation. The FAIR principles for research software (FAIR4RS) \cite{Barker2022-mn} recommend making software findable through repositories and registries, ensuring accessibility through open licensing and documentation, promoting interoperability through standardization and compatibility with other tools, and enabling reusability through versioning and clear documentation of functionality and dependencies. These principles have made a significant impact on policy and publishing, and they have been taken up by the public and private sector as well as spawning an industry of projects, research programs, and commercial businesses \cite{industry-fair, harrow2022}.
 
\section*{FAIR Principles applied to computational workflows}\label{sec:fair-principles-applied}
FAIR4RS defines research software as ``source code files, algorithms, scripts, computational workflows, and executables that were created during the research process or for a research purpose''. Workflows are a specific kind of software where reuse and modification should be inherent in their modularized composition of code and sub-workflows with an explicitly defined control and data flow. Workflow composition can be organized along levels of granularity (e.g. functions, tasks, stages, pipelines), for example similar to libraries in a software stack. The execution of a series of computational code can be difficult to standardize and share across different research environments, and each has its own set of dependencies and requirements, which can complicate the process of ensuring comprehensive FAIRness \cite{De_Visser2023-pr}. Automating scientific experiments using workflows helps mitigate this issue. A key characteristic is the separation of the workflow specification from its execution; the description of the process is a form of data-describing method \cite{Goble2020-sn}. Workflows that are chiefly concerned with the processing and creation of data are typically designed to be tightly coupled with their data, including test data and the provenance lineage history of data products. Therefore, workflows have an important role to play in ensuring that the data products they use and produce are FAIR.

Building on earlier work \cite{Goble2020-sn,wolf2021,Wilkinson2022,De_Visser2023-pr,Niehues2024,Zulfiqar2024}, the WCI FAIR Computational Workflows Working Group (WCI-FW) set out to systematically apply the FAIR principles to computational workflows. WCI-FW is an open, diverse, and interdisciplinary collective of workflow system professionals and workflow users and authors who meet bi-weekly online. The group comprises experts from 40 organizations from 13 countries specialized in various fields, including bioinformatics, climate science, data science, earth science, software engineering, and workflow management. This diversity of backgrounds and expertise allowed us to approach the task from multiple perspectives, ensuring that the FAIR principles for computational workflows are both practical and widely accepted to enhance the reusability, reliability, and impact of computational workflows across diverse research domains.

The WCI-FW debated for many months whether to define a new, tailored set of principles for computational workflows or to apply the established principles for data and software. We concluded that we would directly apply the principles for FAIR data and software after recognizing that (a) we were naturally developing variations of these established principles because (b) workflows are a hybrid form of software, closely bound to data and even possessing data-like properties \cite{Goble2020-sn}. By building on the established FAIR framework, we have maintained consistency and enabled practices for workflows to integrate seamlessly with existing practices for data and software.

Several pivotal questions surfaced in the course of our group discussions:
\begin{itemize}
    \item Do we need to tailor software principles to the specific characteristics of workflows? As workflows are executables, the principles of software should apply. Nevertheless, the characteristics of the heterogeneous components that make up workflows and the expected user behavior of variant reuse, modification, and portability led us to some customization of the principles. Technical information is needed that details each step’s inputs, outputs, dependencies, and computational requirements as well as configuration files, lists of software dependencies, and other information about the operational context. Missing context is one of the main difficulties faced when porting across computing platforms; applying FAIR to workflows requires fully describing the context necessary for executing the workflows as software.
    \item How far do the data principles apply to workflows as digital data objects? Workflows and their components are frequently represented as collections of files, just like other datasets, and these files are interpreted as different kinds of objects like source code, raw data, and AI models. FAIR workflows augment this idea by encouraging comprehensive workflow ``datasets'' to include detailed metadata and documentation to describe their structures, purposes, and technical requirements. We applied the FAIR principles for data to the collection of files that represent the workflow. For example, these files should include descriptive metadata such as title, authors, creation date, and version, as well as version histories when possible. Workflows need persistent identifiers just like other datasets do, and they need to be accessible as data that can be retrieved over open protocols. Workflows also need licenses that explain clearly the conditions under which others are permitted to use and modify parts or the whole.
    \item Given that FAIR is primarily about persistent identifiers and metadata, what should be identified for a workflow, and what is the necessary metadata? Workflows are composite objects, where the entire object (with a persistent identifier and metadata) as well as its components (also with persistent identifiers and metadata), all in turn need to be FAIR. Executable components can be independent of the workflow (call-outs to tools, for example), embedded (internal scripts), or workflows themselves (so that FAIR principles are applied recursively).  Data components may be interpreted as data and as metadata. Workflows provide mechanisms to execute in one run independent components that may be independently FAIR, but their aggregation into a workflow also needs to be FAIR.
\end{itemize}

Table~\ref{tab:1} summarizes our results with further discussion below. In a nutshell, \textbf{Findability} means that a workflow and its associated metadata are easy for both humans and machines to find. \textbf{Accessibility} means that a workflow and its metadata are retrievable via standardized protocols. \textbf{Interoperability} means that a workflow interoperates with other workflows, and workflow components interoperate, by exchanging data and/or metadata, and/or through interaction via APIs, described by standards. Finally, \textbf{Reusability} means that a workflow is both usable (can be executed) and reusable (can be understood, modified, built upon, or incorporated into other workflows).

\begin{table}[ht]\tiny
\caption{The FAIR Principles Applied to Computational Workflows}\label{tab:1}
\begin{tabular}{|p{0.9\textwidth}|p{0.1\textwidth}|}
\hline
Guideline & Based on\footnote{1} \\
\hline
F1. A workflow is assigned a globally unique and persistent identifier.
                                            & D-F1 \newline S-F1\\ \hline & \\
   F1.1. Components of the workflow representing levels of granularity are assigned distinct identifiers.& S-F1.1 \\ \hline & \\
F1.2. Different versions of the workflow are assigned distinct identifiers.
                                                        & S-F1.2 \\ \hline & \\
F2. A workflow and its components are described with rich metadata.
                                            & D-F2 \newline S-F2 \\ \hline & \\
F3. Metadata clearly and explicitly include the identifier of the workflow, and workflow versions, that they describe.
                                            & D-F3 \newline S-F3 \\ \hline & \\
F4. Metadata and workflow are registered or indexed in a searchable FAIR resource.
                                            & D-F4 \newline S-F4 \\ \hline & \\
A1. Workflow and its components are retrievable by their identifiers using a standardized communications protocol.
                                            & D-A1 \newline S-A1 \\ \hline & \\
A1.1. The protocol is open, free, and universally implementable.
                                        & D-A1.1 \newline S-A1.1 \\ \hline & \\
A1.2. The protocol allows for an authentication and authorization procedure, when necessary.
                                        & D-A1.2 \newline S-A1.2 \\ \hline & \\
A2. Metadata are accessible, even when the workflow is no longer available.
                                        & D-A2 \newline S-A2 \\ \hline & \\
I1. Workflow and its metadata (including workflow run provenance) use a formal, accessible, shared, transparent, and broadly applicable language for knowledge representation.
                                        & D-I1 \newline S-R1.2 \\ \hline & \\
I2. Metadata and workflow use vocabularies that follow FAIR principles.
                                        & D-I2 \\ \hline & \\
I3. Workflow is specified in a way that allows its components to read, write, and exchange data (including intermediate data), in a way that meets domain-relevant standards.
                                        & D-R3 \newline S-I1 \\ \hline & \\
I4. Workflow and its metadata (including workflow run provenance) include qualified references to other objects and the workflow's components.
                            & D-I3 \newline S-I2 \newline S-R1.2 \\ \hline & \\
R1. Workflow is described with a plurality of accurate and relevant attributes.
                                & D-R1 \newline S-R1 \\ \hline & \\
R1.1. Workflow is released with a clear and accessible license.
                                & D-R1.1 \newline S-R1.1 \\ \hline & \\
R1.2. Components of the workflow representing levels of granularity are given clear and accessible licenses.
                                & D-R1.1 \newline S-R1.1 \\ \hline & \\
R1.3. Workflow is associated with detailed provenance of the workflow and of the products of the workflow.
                                & D-R1.2 \newline S-R1.2 \\ \hline & \\
R2. Workflow includes qualified references to other workflows.
                                & D-I3 \newline S-R2 \\ \hline & \\
R3. Workflow meets domain-relevant community standards.
                                & D-R1.3 \newline S-R3 \\ \hline
\end{tabular}
\footnotetext[1]{The ``Based on'' column references the source rule from the FAIR principles for [D]ata~\cite{Wilkinson2016-vi} and [S]oftware~\cite{Barker2022-mn}.}
\end{table}

\subsection*{Findability}\label{subsec:findability}

The data and software principles generally apply to workflows, but with novel/additional considerations on our interpretation of a component and a strengthening of metadata associated with workflow versions. Workflows are subject to continuous refinement and independent modification, often resulting in multiple versions, adaptations, and variants, with workflows extended, nested, and reused as sub-workflows. Researchers modify workflows for specific purposes, adjust components, and provide alternative methods across various platforms; components are copied, pasted, and modified.

Identifiers are important because unlike software, authors often neglect to  ``name'' their workflows (although they do name their workflow management systems). Identifiers establish a clear lineage of each workflow and component's origin and changes, enabling researchers to understand the workflow structure and evolution, compare iterations, relate workflows to each other, and confidently build upon previous work. Persistent identifiers identify and reference specific versions to ensure accurate identification and traceability and to support developer citation and source workflow attribution (F1). Unique identifiers to different versions of workflows recognize their dynamic nature in research (F1.2). Unique identifiers for a workflow's components and versions (whether a file, computational step, or sub-workflow) support traceability and facilitate reuse, reproducibility, and portability across different workflow versions (F1.1). 

Rich metadata for the overall workflows and their components (including inputs and outputs) (F2, F3) should use a formal, accessible language for knowledge representation, such as schema.org (\url{https://schema.org}), and standardized metadata such as the Bioschemas schema.org profile for Computational Workflows (\url{https://bioschemas.org/types/ComputationalWorkflow/}), used for workflow components in Workflow Run Crate profile \cite{Leo2024} and CodeMeta (\url{https://w3id.org/codemeta/v3.0}). Other metadata forms for workflows include canonical ``Rosetta'' languages (e.g. CWL \cite{Crusoe2022}, WDL \cite{Voss2017}) that abstract from the native workflow language for greater portability and comparability. Such metadata are used for registration and search in dedicated and trusted workflow registries such as WorkflowHub \cite{Goble2021} and Dockstore (F4) \cite{Yuen2021}.

\subsection*{Accessibility}\label{subsec:accessibility}

The data and software principles generally apply but with new considerations on what it means for a workflow to be available. The workflow may be restricted to execution at one facility, which is a common case in highly optimized HPC settings, and access to that facility may be restricted (A1.2). Single sign-on for workflow components requires harmonized Authentication and Authorization Infrastructure (AAI) propagation through the different tasks, if they are hosted by different service providers. ``A2. Metadata are accessible, even when the workflow is no longer available'' requires further interpretation given a workflow’s dual nature as process description (data) and process execution (software), and we must understand what ``no longer available'' means. As a piece of software, a workflow execution platform or the infrastructure that it operates on may become deprecated and/or unavailable. Dependency on the component code and services makes a workflow vulnerable to ``software rot'', and it is not always possible to completely containerize and preserve all code, such as when they are remotely executed third party tools or call-outs to online datasets \cite{Zhao2012}.

As a process description, the workflow captures both the steps and data flow of a recipe for a method, which is a form of the ``metadata''. Thus at least the workflow and its components descriptions, and if relevant the workflow run components, should be preserved and archived in a long-term registry or repository, so that if the workflow is no longer executable, it will still be readable. If we interpret a workflow description file as data, the data principles decree that at least the metadata for that file should still be retrievable. For a workflow, however, that data file is the essence of the workflow. We interpret the workflow description to be a form of metadata of the software and therefore, it must always be accessible. A workflow associated with a publication may still be examined as a method, and its provenance may be inspected; the description should be enough to be translated for another workflow system. Apart from manual or AI-assisted porting of code, this can also happen via ``Rosetta'' workflow languages like CWL and WDL (as e.g. done in MGnify \cite{richardson2023} and by Snakemake's CWL export).

\subsection*{Interoperability}\label{subsec:interoperability}

The data and software principles are combined and blur with principles from Reusability, as we tease apart the FAIR requirements for a workflow, its components (both executable and data), and an instantiated and executed run of the workflow. For example, principle I1 applies to describing the static workflow structure of the specification and capturing dynamic aspects of the run such as execution details and intermediate outputs. The workflow specification plays a role in data in I1, I2, and I4 adapting data principles, as well as the role of metadata of the software, co-opting software principles.

We adapt software principles for I3 and I4 to explicitly refer to the domain standard description of input and outputs of the workflow executable components. As data components of a workflow, the expected input and output datasets, benchmarks and parameters, and the actual datasets and parameter files from a workflow run, should also adhere to community standards (data principle R1.3), as should intermediate data derived during the execution. The principles encourage workflow portability across different types of computational environments (e.g., via containerization or OS-agnostic package management) \cite{De_Visser2023-pr} and, ideally, easy translation from one language/management system to another or at least compatibility and seamless integration (in case of sub-workflows). 

Standard language formats like CWL encourage interoperability and offer the potential for workflow comparisons and exchange between different languages, through documentation of components, inputs, and outputs. We explicitly add workflow run provenance to interoperability as documented execution steps and data lineage traces are important components that should also be interoperable and machine-actionable \cite{nicolae2023}.

\subsection*{Reusability}\label{subsec:reusability}

Reusability of the workflow as an executable method with modifications to its parameters and input files is an expected and anticipated use of the workflow, tied up in the quality of its documentation and the accessibility of computational infrastructure capable of executing it. 

While previously discussing findability, we also highlighted the reusability of workflows as complex compositions intended to be modified, built upon, and incorporated into other workflows, as well as their reusability as runnable executions with modified datasets and parameter files \cite{Lamprecht2021}. The data and software principles are adapted to emphasize the composite nature of workflows (both executable and data components) for licensing \cite{Cohen-Boulakia2017-pw}. Workflows are composed of many parts, potentially from different authors with different intents for how their artifacts may be used; these intents must be clearly stated by the inclusion of licenses that cover all parts of the workflow and sub-parts of the workflows that can be reused as sub-workflows \cite{De_Visser2023-pr}.

The provenance history of the workflow – its authors, purpose, workflows of which it may be a variant, links to other sub-workflows and so on, are included in R1.3 drawing from both data and software principles R1.2. The provenance of data products tracked by the workflow run provenance collection is related to the data principles R1.2 where data (in this case the products of the workflow) are associated with detailed provenance that a workflow system should be able to provide. A benefit of using fully fledged workflow systems is the capture of computer-interpretable provenance (meta)data to track the history and origin of data processing steps. R1 and R3 expect that input and output datasets are thoroughly described with example input and expected output data for each step that adhere to data structure and file format standards \cite{De_Visser2023-pr} and gold standards for benchmarking \cite{Cohen-Boulakia2017-pw}.

\section*{Examples}\label{sec:examples}
Here, we provide concrete examples with detailed descriptions for applying the FAIR principles to both workflow specifications and workflow runs. For each case, we narrate through the FAIR principles sequentially. We recommend the reader to follow along with Table~\ref{tab:1} as a guide.

\subsection*{Example workflow specification}\label{subsec:example-workflow-specification}

The Protein MD Setup workflow sets up a protein molecular dynamic simulation that returns a protein structure and simulated 3D trajectories \cite{SoilandReyes2022}. It is registered in WorkflowHub (\url{https://doi.org/10.48546/workflowhub.workflow.29.3}), where it has been assigned a Digital Object Identifier (DOI) which is globally unique (F1). Different versions of this workflow can be identified using unique digital object identifiers (F1.2). The workflow is described using (rich) metadata (F2) that includes the identifier of the workflow and its unique version (F3). The workflow and associated metadata are registered (F4). Moreover, components of the workflow have their own individual persistent identifiers as the workflow is composed of sub-workflow BioBB building blocks from the BioBB Library (\url{https://workflowhub.eu/projects/11}) (F1.2). The workflow and its components can be retrieved and downloaded using a standardized HTTPS protocol (A1, A1.1, A1.2). The metadata associated with the workflow and its components is independent from the GitHub code repository and are accessible on WorkflowHub even if the workflow (or its components) become inaccessible (A2).

The Protein MD Setup is written using CWL which follows a formal and declarative paradigm for workflow implementation (I1). The workflow does not include all the best-practice principles to fulfill I2 and I3. This extends to the workflow metadata, which does not use a domain-specific ontology. The workflow and its metadata include qualified references to other objects and components (I4). However, CWL standard itself allows metadata and workflow vocabularies (e.g. Bioschemas, EDAM~\cite{edam2013}) that are consistent with FAIR principles (I2), enables the components of the workflows to read, write, and exchange data using format that meets domain-specific standards (I3), and assigns qualified references to workflow components, inputs, and outputs. 

The workflow is released under Apache License 2.0 software license (R1.1) and  WorkflowHub’s RO-Crate references that the embedded CWL code is archived from the corresponding GitHub source code repository (\url{https://github.com/bioexcel/biobb-wf-md-setup-protein-cwl}). The component BioBB building blocks of the workflow have their own clear and accessible licenses (R1.2). The workflow is associated with detailed provenance about its development, provided as a downloadable RO-Crate object (R1.3).

\subsection*{Example workflow run}\label{subsec:example-workflow-run}

Here, we demonstrate the application of the FAIR principles to a workflow run. Recall from our previous definitions that a workflow run is the instantiation of a workflow with inputs (parameters files, input datasets) and outputs (output data, the provenance execution log and lineage of data products). In this example, the workflow run is instantiated from a tissue/tumor prediction workflow specification for digital pathology, and it is represented using RO-Crate \cite{leo2023}.

The workflow run has been assigned a globally unique DOI (F1), and components of the workflow run (e.g. steps, inputs, outputs) are also assigned their own distinct identifiers (F1.1). The workflow run is described using rich metadata (F2) that includes the identifier of the workflow run and its unique version (F3). The corresponding workflow specification is not registered, but it is publicly available via GitHub (\url{https://github.com/crs4/deephealth-pipelines}), and the workflow run is registered on Zenodo, which is a searchable FAIR resource (F4).

The workflow run and its components are also hosted on Zenodo, which means that they can be retrieved and downloaded using the HTTPS protocol, which is standardized (A1), open and free and universally implementable (A1.1), and able to support authentication and authorization procedures (A1.2). The metadata associated with the workflow run and its components are independent from the workflow run and components themselves, and they are accessible on Zenodo even if the workflow run and its components become inaccessible (A2).

Recall from Figure~\ref{fig:1} that execution provenance is a product of a workflow run. Here, provenance from the digital pathology tissue/tumor prediction workflow run is generated using CWLProv \cite{Khan2019} in the form of RO-Bundle and then converted to an RO-Crate \cite{Sefton2023} that follows the Provenance Run Crate profile (\url{https://www.researchobject.org/workflow-run-crate/profiles/0.1/provenance_run_crate}). All of these methods represent workflow run information by utilizing well-established standards and language (I1). CWLProv and RO-Crate allow vocabularies that are consistent with the FAIR principles (I2). The workflow run provenance is generated using CWLProv and documented as a Provenance Run Crate profile showing the exchange of data between components of the workflow.  The workflow run does not, however, include domain-specific standards for metadata to fulfill (which means it does not satisfy I3). The workflow run provenance does include qualified references to other objects and the workflow's components (I4).

The workflow run is described with many accurate and relevant attributes (R1), and it is released under the clear and accessible MIT License (R1.1). One of the workflow run's components, Slaid, is a library for applying DL models from the DeepHealth project (\url{https://deephealth-project.eu/}) on whole slide images (WSI) that is also released independently under the MIT License (R1.2). The workflow run contains detailed provenance about the execution, provided as a downloadable RO-Crate object (R1.3). The workflow run processes digital pathology images in the formats supported by OpenSlide (R3), which is a widespread software library that represents a community standard for digital pathology images \cite{openslide2013}.
 
\section*{FAIRness beyond the FAIR Principles}\label{sec:beyond-fairness}
The FAIR principles in Table~\ref{tab:1} focus on the workflows as scholarly research assets to be found, accessed, interoperated, and reused. They do not extend to the quality of the workflows (FAIR+Q) or the extent of the reproducibility of the workflows (FAIR+R), and they only hint at their portability and sustainability. The Open Data, Open Code, Open Infrastructure (O3) guidelines complement FAIR by providing an actionable road map toward sustainability \cite{Hoyt2024}. Note, however, that FAIR does not require openness; authors and companies may even choose not to publish their workflows or metadata at all, but still use ``inner FAIR'' principles \cite{CrusoeTwitter} for their own benefits.

Workflows are composites of executable components that need to be validated and tested with clean interfaces, permissive access permissions if remotely executed, and compatible licenses. Work on canonical workflow libraries \cite{Turilli2019} and building blocks that are interoperable between languages \cite{SoilandReyes2022}, workflow readiness of tools \cite{Brack2022}, FAIR unit testing like pytest-workflow for WDL (\url{https://github.com/LUMC/pytest-workflow}), test monitoring like LifeMonitor(\url{https://crs4.github.io/life_monitor/}), and benchmarking like OpenEBench (\url{https://openebench.bsc.es/}) are steps towards creating FAIR workflow professional practices that should include peer review, curation, and perhaps even certification.

We should also consider the relationship of FAIR to other services in a workflow's service ecosystem. Container technologies such as Docker and Singularity play a role in the interoperability, reusability, and reproducibility of computational workflows by providing lightweight, portable, and self-contained environments that include all of the software dependencies and libraries required to run a workflow. In alignment with the FAIR principles, containers can be versioned and shared through registries like Docker Hub, Nexus Repository, or Github Container registry, but these registries are not truly FAIR because old versions may not be preserved. FAIR workflows should be stored in version-controlled repositories like GitHub and GitLab and indexed by registries like Dockstore \cite{Yuen2021} and WorkflowHub \cite{Goble2021}. Workflow registries like Dockstore and WorkflowHub support findability and accessibility so workflows can be shared, published, and reused as well as downloaded or launched using dedicated infrastructure facilities such as Galaxy Europe. Both support example input and test data components for a workflow specification but do not retain the result components of a workflow run -- the provenance and links to datasets that are the resulting products. That is expected to be managed by data repositories like Zenodo and Figshare that in turn need to support FAIR data principles. However, this container approach most probably shows practical limitations when applied to workflows executed on very large-scale, meticulously maintained and pampered HPC environments running on the verge of computational and technical stability. For such cases, reproducibility of workflows might always be an issue. 

Workflows are multi-part objects whose data components are as important as their executable steps; all these digital objects should be FAIR too. Community efforts like RO-Crate \cite{SoilandReyes2022ROcrate} propose the means to encapsulate workflows and all their components while capturing the context in which they are used (e.g., executions, bibliography, sketches, etc.). An RO-Crate Workflow profile (\url{https://about.workflowhub.eu/Workflow-RO-Crate/}) includes references to components which also need to be FAIR. By having workflows as composition of FAIR research outputs, the FAIRness of each contained resource is validated, since the execution of the workflow relies on these resources being available. Services like WorkflowHub (registry), LifeMonitor (test monitoring), and Galaxy (workflow platform) implement RO-Crate for comprehensive workflow representation and exchange as well as a step to sharing metadata and persistent identifiers to implement the FAIR principles of Table~\ref{tab:1}.

As workflows typically operate on and produce datasets, they are well placed to make those data products FAIR; they provide the process description and automated data provenance collection, and fully fledged WMS can automate processes for generating FAIR data. Nevertheless, workflows have to be well designed to be ``FAIR aware'': producing and consuming data in standardized formats, assigning license, handling identifiers, and metadata production, data versioning, etc \cite{Goble2020-sn}. Using workflows as automated instruments for ``FAIR data by design'' can encourage inputs and make outputs to be machine-actionable and more suitable for use in other workflows, for example those that span geographically distinct computing facilities \cite{Antypas2021-ph}.
 
\section*{Conclusion}\label{sec:conclusion}
Computational workflows are key instruments for the advancement in scientific research, potentially revolutionizing how data is managed, analyzed, and shared. Workflows not only serve as software that can be used by experts and non-coders alike; they also serve as documentation of scientific processes, encapsulating not only the data and software used but also the context and conditions under which discoveries were made. It is expected that this comprehensive documentation will not only enhance the credibility of research findings but will also facilitate the replication and validation of scientific results across diverse settings and by different research teams.

FAIR data and FAIR software principles both apply to workflows, combining structured, well-described data with portable, well-documented software creating a powerful framework for advancing scientific knowledge. Moreover, the next generation of workflows -- assisted auto-assembly, generative workflows, dynamic reconfiguration, and so on -- will require rich, machine-actionable metadata.

FAIR application requires interpretation, however, to consider the abstraction and the compositional properties of workflows. The usual challenges for FAIR implementation are still present: incentives, resources, assessment, support services, and so on. Even when opportunities arise for metadata enrichment when registering workflows in WorkflowHub, for example, the best curators do not manage all the principles, and most take the simplest route. FAIR workflow implementation will ``take a village'' of services, including data services and support from workflow management systems, and well-designed workflows that are FAIR data-aware will require golden standard examples, training, and professionalization.
 
\backmatter

\bmhead{Acknowledgements}
This research used resources of: the Oak Ridge Leadership Computing Facility at the Oak Ridge National Laboratory, which is supported by the Office of Science of the U.S. Department of Energy under Contract No. DE-AC05-00OR22725 (SRW); Sandia National Laboratories, a multi-mission laboratory managed and operated by National Technology \& Engineering Solutions of Sandia, LLC (NTESS), a wholly owned subsidiary of Honeywell International Inc., for the U.S. Department of Energy's National Nuclear Security Administration (DOE/NNSA) under contract DE-NA0003525 (LP); the European Union programme Horizon Europe under grant agreements HORIZON-INFRA-2021-EOSC-01 101057388 (EuroScienceGateway),  HORIZON-INFRA-2023-EOSC-01-02 101129744 (EVERSE; SS), HORIZON-INFRA-2021-EOSC-01-05 101057344 and by UK Research and Innovation (UKRI) under the UK government's Horizon Europe funding guarantee grants 10038963 (EuroScienceGateway; CG, S-SR), 10038992 (FAIR-IMPACT; NJ); Australian BioCommons, which is enabled by NCRIS via Bioplatforms Australia funding (JG); Deutsche Forschungsgemeinschaft (DFG, German Research Foundation) as part of GHGA – The German Human Genome-Phenome Archive (www.ghga.de, Grant Number 441914366 (NFDI 1/1)); the National Research Agency under the France 2030 program, with reference to ANR-22-PESN0007 (KB).
 
\bmhead{Author Contributions}

SRW and CG are the primary authors of the manuscript. All other authors are listed alphabetically and contributed to the manuscript by participating in the WCI-FW meetings and by editing or commenting on the manuscript text.

\bmhead{Competing Interests}

The authors declare no competing interests.


\begin{thebibliography}{10}
\expandafter\ifx\csname url\endcsname\relax
  \def\url#1{\burl{#1}}\fi
\expandafter\ifx\csname urlprefix\endcsname\relax\def\urlprefix{URL }\fi
\providecommand{\bibinfo}[2]{#2}
\providecommand{\eprint}[2][]{\url{#2}}
\providecommand{\doi}[1]{\url{https://doi.org/#1}}
\bibcommenthead

\bibitem{SCHINTKE202482}
\bibinfo{author}{Schintke, F.} \emph{et~al.}
\newblock \bibinfo{title}{Validity constraints for data analysis workflows}.
\newblock \emph{\bibinfo{journal}{Future Generation Computer Systems}}
  \textbf{\bibinfo{volume}{157}}, \bibinfo{pages}{82--97}
  (\bibinfo{year}{2024}).
\newblock \urlprefix\url{https://doi.org/10.1016/j.future.2024.03.037}.

\bibitem{DiTommaso2017}
\bibinfo{author}{Di~Tommaso, P.} \emph{et~al.}
\newblock \bibinfo{title}{Nextflow enables reproducible computational
  workflows}.
\newblock \emph{\bibinfo{journal}{Nature Biotechnology}}
  \textbf{\bibinfo{volume}{35}}, \bibinfo{pages}{316–319}
  (\bibinfo{year}{2017}).
\newblock \urlprefix\url{https://doi.org/10.1038/nbt.3820}.

\bibitem{Galaxy2024}
\bibinfo{author}{Abueg, L. A.~L.} \emph{et~al.}
\newblock \bibinfo{title}{The galaxy platform for accessible, reproducible, and
  collaborative data analyses: 2024 update}.
\newblock \emph{\bibinfo{journal}{Nucleic Acids Research}}
  \textbf{\bibinfo{volume}{52}}, \bibinfo{pages}{W83–W94}
  (\bibinfo{year}{2024}).
\newblock \urlprefix\url{https://doi.org/10.1093/nar/gkae410}.

\bibitem{Molder2021}
\bibinfo{author}{M\"{o}lder, F.} \emph{et~al.}
\newblock \bibinfo{title}{Sustainable data analysis with snakemake}.
\newblock \emph{\bibinfo{journal}{F1000Research}}
  \textbf{\bibinfo{volume}{10}}, \bibinfo{pages}{33} (\bibinfo{year}{2021}).
\newblock \urlprefix\url{https://doi.org/10.12688/f1000research.29032.2}.

\bibitem{Babuji2019}
\bibinfo{author}{Babuji, Y.} \emph{et~al.}
\newblock \emph{\bibinfo{title}{Parsl: Pervasive parallel programming in
  python}}, HPDC '19 (\bibinfo{publisher}{ACM}, \bibinfo{year}{2019}).
\newblock \urlprefix\url{https://doi.org/10.1145/3307681.3325400}.

\bibitem{kanwal-2017}
\bibinfo{author}{Kanwal, S.}, \bibinfo{author}{Khan, F.~Z.},
  \bibinfo{author}{Lonie, A.} \& \bibinfo{author}{Sinnott, R.~O.}
\newblock \bibinfo{title}{Investigating reproducibility and tracking provenance
  --a genomic workflow case study}.
\newblock \emph{\bibinfo{journal}{BMC Bioinformatics}}
  \textbf{\bibinfo{volume}{18}}, \bibinfo{pages}{337} (\bibinfo{year}{2017}).
\newblock \urlprefix\url{https://doi.org/10.1186/s12859-017-1747-0}.

\bibitem{roadmap2022}
\bibinfo{author}{Ferreira~da Silva, R.} \emph{et~al.}
\newblock \bibinfo{title}{Workflows community summit 2022: A roadmap
  revolution} (\bibinfo{year}{2023}).
\newblock \urlprefix\url{https://doi.org/10.5281/zenodo.7750670}.

\bibitem{garijo2014}
\bibinfo{author}{Garijo, D.} \emph{et~al.}
\newblock \emph{\bibinfo{title}{Workflow reuse in practice: A study of
  neuroimaging pipeline users}}, Vol.~\bibinfo{volume}{1},
  \bibinfo{pages}{239--246} (\bibinfo{year}{2014}).
\newblock \urlprefix\url{https://doi.org/10.1109/eScience.2014.33}.

\bibitem{Wilkinson2016-vi}
\bibinfo{author}{Wilkinson, M.~D.} \emph{et~al.}
\newblock \bibinfo{title}{The {FAIR} guiding principles for scientific data
  management and stewardship}.
\newblock \emph{\bibinfo{journal}{Sci. Data}} \textbf{\bibinfo{volume}{3}},
  \bibinfo{pages}{160018} (\bibinfo{year}{2016}).
\newblock \urlprefix\url{https://doi.org/10.1038/sdata.2016.18}.

\bibitem{Barker2022-mn}
\bibinfo{author}{Barker, M.} \emph{et~al.}
\newblock \bibinfo{title}{Introducing the {FAIR} principles for research
  software}.
\newblock \emph{\bibinfo{journal}{Sci. Data}} \textbf{\bibinfo{volume}{9}},
  \bibinfo{pages}{622} (\bibinfo{year}{2022}).
\newblock \urlprefix\url{https://doi.org/10.1038/s41597-022-01710-x}.

\bibitem{industry-fair}
\bibinfo{author}{van Vlijmen, H.} \emph{et~al.}
\newblock \bibinfo{title}{The need of industry to go fair}.
\newblock \emph{\bibinfo{journal}{Data Intelligence}}
  \textbf{\bibinfo{volume}{2}}, \bibinfo{pages}{276--284}
  (\bibinfo{year}{2020}).
\newblock \urlprefix\url{https://doi.org/10.1162/dint\_a\_00050}.

\bibitem{harrow2022}
\bibinfo{author}{Harrow, I.}, \bibinfo{author}{Balakrishnan, R.},
  \bibinfo{author}{{Küçük McGinty}, H.}, \bibinfo{author}{Plasterer, T.} \&
  \bibinfo{author}{Romacker, M.}
\newblock \bibinfo{title}{Maximizing data value for biopharma through fair and
  quality implementation: Fair plus q}.
\newblock \emph{\bibinfo{journal}{Drug Discovery Today}}
  \textbf{\bibinfo{volume}{27}}, \bibinfo{pages}{1441--1447}
  (\bibinfo{year}{2022}).
\newblock \urlprefix\url{https://doi.org/10.1016/j.drudis.2022.01.006}.

\bibitem{De_Visser2023-pr}
\bibinfo{author}{de~Visser, C.} \emph{et~al.}
\newblock \bibinfo{title}{Ten quick tips for building {FAIR} workflows}.
\newblock \emph{\bibinfo{journal}{PLoS Comput. Biol.}}
  \textbf{\bibinfo{volume}{19}}, \bibinfo{pages}{e1011369}
  (\bibinfo{year}{2023}).
\newblock \urlprefix\url{https://doi.org/10.1371/journal.pcbi.1011369}.

\bibitem{Goble2020-sn}
\bibinfo{author}{Goble, C.} \emph{et~al.}
\newblock \bibinfo{title}{{FAIR} computational workflows}.
\newblock \emph{\bibinfo{journal}{Data Intell.}} \textbf{\bibinfo{volume}{2}},
  \bibinfo{pages}{108--121} (\bibinfo{year}{2020}).
\newblock \urlprefix\url{https://doi.org/10.1162/dint_a_00033}.

\bibitem{wolf2021}
\bibinfo{author}{Wolf, M.} \emph{et~al.}
\newblock \emph{\bibinfo{title}{Reusability first: Toward fair workflows}},
  \bibinfo{pages}{444--455} (\bibinfo{year}{2021}).
\newblock \urlprefix\url{https://doi.org/10.1109/Cluster48925.2021.00053}.

\bibitem{Wilkinson2022}
\bibinfo{author}{Wilkinson, S.~R.} \emph{et~al.}
\newblock \emph{\bibinfo{title}{{F*** workflows: when parts of FAIR are
  missing}}}, \bibinfo{pages}{507–512} (\bibinfo{publisher}{IEEE},
  \bibinfo{year}{2022}).
\newblock \urlprefix\url{https://doi.org/10.1109/eScience55777.2022.00090}.

\bibitem{Niehues2024}
\bibinfo{author}{Niehues, A.} \emph{et~al.}
\newblock \bibinfo{title}{A multi-omics data analysis workflow packaged as a
  fair digital object}.
\newblock \emph{\bibinfo{journal}{GigaScience}} \textbf{\bibinfo{volume}{13}}
  (\bibinfo{year}{2024}).
\newblock \urlprefix\url{https://doi.org/10.1093/gigascience/giad115}.

\bibitem{Zulfiqar2024}
\bibinfo{author}{Zulfiqar, M.} \emph{et~al.}
\newblock \bibinfo{title}{Implementation of fair practices in computational
  metabolomics workflows—a case study}.
\newblock \emph{\bibinfo{journal}{Metabolites}} \textbf{\bibinfo{volume}{14}},
  \bibinfo{pages}{118} (\bibinfo{year}{2024}).
\newblock \urlprefix\url{https://doi.org/10.3390/metabo14020118}.

\bibitem{Leo2024}
\bibinfo{author}{Leo, S.} \emph{et~al.}
\newblock \bibinfo{title}{Recording provenance of workflow runs with ro-crate}.
\newblock \emph{\bibinfo{journal}{PLOS ONE}} \textbf{\bibinfo{volume}{19}},
  \bibinfo{pages}{e0309210} (\bibinfo{year}{2024}).
\newblock \urlprefix\url{https://doi.org/10.1371/journal.pone.0309210}.

\bibitem{Crusoe2022}
\bibinfo{author}{Crusoe, M.~R.} \emph{et~al.}
\newblock \bibinfo{title}{Methods included: standardizing computational reuse
  and portability with the common workflow language}.
\newblock \emph{\bibinfo{journal}{Communications of the ACM}}
  \textbf{\bibinfo{volume}{65}}, \bibinfo{pages}{54–63}
  (\bibinfo{year}{2022}).
\newblock \urlprefix\url{https://doi.org/10.1145/3486897}.

\bibitem{Voss2017}
\bibinfo{author}{Voss, K.}, \bibinfo{author}{Auwera, G. V.~D.} \&
  \bibinfo{author}{Gentry, J.}
\newblock \bibinfo{title}{Full-stack genomics pipelining with gatk4 + wdl +
  cromwell} (\bibinfo{year}{2017}).
\newblock \urlprefix\url{https://doi.org/10.7490/f1000research.1114634.1}.

\bibitem{Goble2021}
\bibinfo{author}{Goble, C.} \emph{et~al.}
\newblock \bibinfo{title}{Implementing fair digital objects in the eosc-life
  workflow collaboratory}  (\bibinfo{year}{2021}).
\newblock \urlprefix\url{https://doi.org/10.5281/zenodo.4605654}.

\bibitem{Yuen2021}
\bibinfo{author}{Yuen, D.} \emph{et~al.}
\newblock \bibinfo{title}{The dockstore: enhancing a community platform for
  sharing reproducible and accessible computational protocols}.
\newblock \emph{\bibinfo{journal}{Nucleic Acids Research}}
  \textbf{\bibinfo{volume}{49}}, \bibinfo{pages}{W624–W632}
  (\bibinfo{year}{2021}).
\newblock \urlprefix\url{https://doi.org/10.1093/nar/gkab346}.

\bibitem{Zhao2012}
\bibinfo{author}{Zhao, J.} \emph{et~al.}
\newblock \emph{\bibinfo{title}{Why workflows break -- understanding and
  combating decay in taverna workflows}} (\bibinfo{publisher}{IEEE},
  \bibinfo{year}{2012}).
\newblock \urlprefix\url{https://doi.org/10.1109/eScience.2012.6404482}.

\bibitem{richardson2023}
\bibinfo{author}{Richardson, L.} \emph{et~al.}
\newblock \bibinfo{title}{{MGnify: the microbiome sequence data analysis
  resource in 2023}}.
\newblock \emph{\bibinfo{journal}{Nucleic Acids Research}}
  \textbf{\bibinfo{volume}{51}}, \bibinfo{pages}{D753--D759}
  (\bibinfo{year}{2022}).
\newblock \urlprefix\url{https://doi.org/10.1093/nar/gkac1080}.

\bibitem{nicolae2023}
\bibinfo{author}{Nicolae, B.} \emph{et~al.}
\newblock \emph{\bibinfo{title}{Building the i (interoperability) of fair for
  performance reproducibility of large-scale composable workflows in recup}},
  \bibinfo{pages}{1--7} (\bibinfo{year}{2023}).
\newblock \urlprefix\url{https://doi.org/10.1109/e-Science58273.2023.10254808}.

\bibitem{Lamprecht2021}
\bibinfo{author}{Lamprecht, A.-L.} \emph{et~al.}
\newblock \bibinfo{title}{Perspectives on automated composition of workflows in
  the life sciences}.
\newblock \emph{\bibinfo{journal}{F1000Research}}
  \textbf{\bibinfo{volume}{10}}, \bibinfo{pages}{897} (\bibinfo{year}{2021}).
\newblock \urlprefix\url{https://doi.org/10.12688/f1000research.54159.1}.

\bibitem{Cohen-Boulakia2017-pw}
\bibinfo{author}{Cohen-Boulakia, S.} \emph{et~al.}
\newblock \bibinfo{title}{Scientific workflows for computational
  reproducibility in the life sciences: Status, challenges and opportunities}.
\newblock \emph{\bibinfo{journal}{Future Gener. Comput. Syst.}}
  \textbf{\bibinfo{volume}{75}}, \bibinfo{pages}{284--298}
  (\bibinfo{year}{2017}).
\newblock \urlprefix\url{https://doi.org/10.1016/j.future.2017.01.012}.

\bibitem{SoilandReyes2022}
\bibinfo{author}{Soiland-Reyes, S.} \emph{et~al.}
\newblock \bibinfo{title}{Making canonical workflow building blocks
  interoperable across workflow languages}.
\newblock \emph{\bibinfo{journal}{Data Intelligence}}
  \textbf{\bibinfo{volume}{4}}, \bibinfo{pages}{342–357}
  (\bibinfo{year}{2022}).
\newblock \urlprefix\url{https://doi.org/10.1162/dint_a_00135}.

\bibitem{edam2013}
\bibinfo{author}{Ison, J.} \emph{et~al.}
\newblock \bibinfo{title}{{EDAM: an ontology of bioinformatics operations,
  types of data and identifiers, topics and formats}}.
\newblock \emph{\bibinfo{journal}{Bioinformatics}}
  \textbf{\bibinfo{volume}{29}}, \bibinfo{pages}{1325--1332}
  (\bibinfo{year}{2013}).
\newblock \urlprefix\url{https://doi.org/10.1093/bioinformatics/btt113}.

\bibitem{leo2023}
\bibinfo{author}{Leo, S.}
\newblock \bibinfo{title}{Run of digital pathology tissue/tumor prediction
  workflow} (\bibinfo{year}{2023}).
\newblock \urlprefix\url{https://doi.org/10.5281/zenodo.7774351}.

\bibitem{Khan2019}
\bibinfo{author}{Khan, F.~Z.} \emph{et~al.}
\newblock \bibinfo{title}{Sharing interoperable workflow provenance: A review
  of best practices and their practical application in cwlprov}.
\newblock \emph{\bibinfo{journal}{GigaScience}} \textbf{\bibinfo{volume}{8}}
  (\bibinfo{year}{2019}).
\newblock \urlprefix\url{https://doi.org/10.1093/gigascience/giz095}.

\bibitem{Sefton2023}
\bibinfo{author}{Sefton, P.} \emph{et~al.}
\newblock \bibinfo{title}{Ro-crate metadata specification 1.1.3}
  (\bibinfo{year}{2023}).
\newblock \urlprefix\url{https://doi.org/10.5281/zenodo.7867028}.

\bibitem{openslide2013}
\bibinfo{author}{Goode, A.}, \bibinfo{author}{Gilbert, B.},
  \bibinfo{author}{Harkes, J.}, \bibinfo{author}{Jukic, D.} \&
  \bibinfo{author}{Satyanarayanan, M.}
\newblock \bibinfo{title}{Openslide: A vendor-neutral software foundation for
  digital pathology}.
\newblock \emph{\bibinfo{journal}{Journal of Pathology Informatics}}
  \textbf{\bibinfo{volume}{4}}, \bibinfo{pages}{27} (\bibinfo{year}{2013}).
\newblock \urlprefix\url{https://doi.org/10.4103/2153-3539.119005}.

\bibitem{Hoyt2024}
\bibinfo{author}{Hoyt, C.~T.} \& \bibinfo{author}{Gyori, B.~M.}
\newblock \bibinfo{title}{The o3 guidelines: open data, open code, and open
  infrastructure for sustainable curated scientific resources}.
\newblock \emph{\bibinfo{journal}{Scientific Data}}
  \textbf{\bibinfo{volume}{11}} (\bibinfo{year}{2024}).
\newblock \urlprefix\url{https://doi.org/10.1038/s41597-024-03406-w}.

\bibitem{CrusoeTwitter}
\bibinfo{title}{From a chat with @soilandreyes, the idea of ``inner fair''', in
  the spirit of ``inner source'''}.
\newblock
  \bibinfo{howpublished}{\url{https://web.archive.org/web/20221224155747/https://twitter.com/biocrusoe/status/976827491460493312}}.
\newblock \bibinfo{note}{Published: Twitter 2018-02-22, Archived: 2024-12-24}.

\bibitem{Turilli2019}
\bibinfo{author}{Turilli, M.}, \bibinfo{author}{Balasubramanian, V.},
  \bibinfo{author}{Merzky, A.}, \bibinfo{author}{Paraskevakos, I.} \&
  \bibinfo{author}{Jha, S.}
\newblock \bibinfo{title}{Middleware building blocks for workflow systems}.
\newblock \emph{\bibinfo{journal}{Computing in Science \& Engineering}}
  \textbf{\bibinfo{volume}{21}}, \bibinfo{pages}{62–75}
  (\bibinfo{year}{2019}).
\newblock \urlprefix\url{https://doi.org/10.1109/MCSE.2019.2920048}.

\bibitem{Brack2022}
\bibinfo{author}{Brack, P.} \emph{et~al.}
\newblock \bibinfo{title}{Ten simple rules for making a software tool
  workflow-ready}.
\newblock \emph{\bibinfo{journal}{PLOS Computational Biology}}
  \textbf{\bibinfo{volume}{18}}, \bibinfo{pages}{e1009823}
  (\bibinfo{year}{2022}).
\newblock \urlprefix\url{https://doi.org/10.1371/journal.pcbi.1009823}.

\bibitem{SoilandReyes2022ROcrate}
\bibinfo{author}{Soiland-Reyes, S.} \emph{et~al.}
\newblock \bibinfo{title}{Packaging research artefacts with ro-crate}.
\newblock \emph{\bibinfo{journal}{Data Science}} \textbf{\bibinfo{volume}{5}},
  \bibinfo{pages}{97–138} (\bibinfo{year}{2022}).
\newblock \urlprefix\url{https://doi.org/10.3233/DS-210053}.

\bibitem{Antypas2021-ph}
\bibinfo{author}{Antypas, K.~B.} \emph{et~al.}
\newblock \emph{\bibinfo{title}{Enabling discovery data science through
  cross-facility workflows}} (\bibinfo{publisher}{IEEE}, \bibinfo{year}{2021}).
\newblock \urlprefix\url{https://doi.org/10.1109/BigData52589.2021.9671421}.

\end{thebibliography}
\end{document}